\documentstyle[12pt]{article}

\begin{document}
\font\titfont=cmbx10 scaled \magstep2
\font\autor=cmr9 scaled\magstep1
\font\autofont=cmbx6 scaled \magstep1
\def\qed{\vrule height 1.2ex width 1.1ex depth -.1ex}
\def\litem{\par\noindent
               \hangindent=\parindent\ltextindent}
\def\ltextindent#1{\hbox to\hangindent{#1\hss}\ignorespaces}
\newcommand{\nz}{\hfill\break\noindent}
\newcommand{\sn}{\smallskip\noindent}
\newcommand\mn{\medskip\noindent}
\newcommand{\bn}{\bigskip\noindent}
\newcommand{\D}{{\cal D}}

\noindent
\centerline{{\titfont Chiral symmetry breaking in gauged ${\bf NJL}$ model }} 
\sn
\centerline{{\titfont in curved spacetime}}
\vspace{1.0cm}

\mn
{\autor\centerline{B. Geyer\thanks{E-mail: geyer@ntz.uni-leipzig.d400.de}}}
{\autor\centerline{Institute of Theoretical Physics and}}
{\autor\centerline{Center for Theoretical Sciences}}
{\autor\centerline{Leipzig University}}
{\autor\centerline{Augustusplatz 10, D-04109 Leipzig, Germany}}
\sn
{\autor\centerline{and}}
{\autor\centerline{S. D. Odintsov\thanks{E-mail: sergei@ecm.ub.es
\protect\newline
On leave from Tomsk Pedagogical University, 634041 Tomsk, Russia.}}}
{\autor\centerline{Dept. ECM, Fac. de Fisica,}}
{\autor\centerline{Universidad de Barcelona,}}
{\autor\centerline{Diagonal 647, 08028 Barcelona, Spain}}

\vspace{1.0cm}

\begin{abstract}

Using the renormalization group (RG) approach and the equivalency between 
the class of gauge-Higgs-Yukawa models and the gauged Nambu-Jona-Lasinio
(NJL) model, we study the gauged NJL model in curved space-time. The 
behaviour of the scalar-gravitational coupling constant $\xi(t)$ in 
both theories is discussed. The RG improved effective potential of gauged 
NJL model in curved spacetime is found. The curvature at which chiral 
symmetry in the gauged NJL model is broken is obtained explicitly in
a remarkably simple form. The powerful RG improved effective potential
formalizm leads to the same results as ladder Schwinger-Dyson equations
which have not been formulated yet in curved spacetime what opens
new possibilities in the study of GUTs and NJL-like models in curved
spacetime.

\vspace{0.5cm}


\vspace{.5cm}
{\footnotesize
\noindent 1~~ E-mail: geyer@ntz.uni-leipzig.d400.de} \\
\noindent 2~~ E-mail: sergei@ecm.ub.es \\\mbox{}~~~ On leave from Tomsk 
Pedagogical University,
634041 Tomsk, Russia\\

\end {abstract}

\newpage

\section{\large\bf Introduction}

It is quite well known that dynamical symmetry breaking in the 
Standard Model (SM) may be realised via Nambu--Jona-Lasinio (NJL)
mechanism \cite{1}, for example, with the help of  quark condensate
(see \cite{2}, \cite{2a}, \cite{3} and references therein.)
In such an approach -- having in mind that in studying composite bound 
states the NJL model can be treated analytically -- it is not by no means
necessary to introduce elementary Higgs fields to discuss dynamical 
symmetry breaking.

A convenient way to study chiral symmetry breaking in the gauged NJL 
model is via Schwinger--Dyson (SD) equations, or equivalently, using 
the effective action for the composite fields \cite{4},\cite{5}. Very often it is 
necessary to study chiral symmetry breaking under some external 
conditions, like non--zero temperature, external electromagnetic or 
gravitational field. For example, the influence of the gravitational 
field on the chiral symmetry breaking may be important in the study of 
problems related to the early universe or to the physics of black holes.

In the past the study of the non-gauged NJL model in curved spacetime (in
linear
curvature approximation for which curvature effects actually may be 
relevant in GUT's and in SM) has shown  quite rich phase
structure and the possibility of curvature--induced transitions \cite{6} 
between chiral symmetric and non--symmetric phases. Such a study has 
been done using the effective potential in the leading order of 
$1/N$-expansion. Now, it is quite interesting to discuss similar 
questions for the more realistic gauged NJL model. This model recently 
has been studied in full detail using the ladder SD equations (see
\cite{7} where also renormalizability of gauged NJL model has been
discussed).
Unfortunately -- due to the absence of the momentum representation in 
general curved spacetime -- all attempts to generalize the SD equations 
to curved spacetime were not quite successful \cite{8}.

In the paper \cite{3} (see also \cite{9}) the NJL model has been discussed using the
standard RG equations for coupling constants with appropriate compositeness
 conditions. This idea has been further elaborated in Ref. \cite{10} for the 
gauged NJL model in flat spacetime. In particular, it has been shown that

some class of
gauge Higgs--Yukawa models in leading order of a modified $1/N_c$ expansion 
leads to a well--defined, non--trivial theory. This theory is equivalent to 
the gauged NJL model when the ultraviolett cutoff goes to infinity (using 
corresponding compositeness conditions).

In the present work we extend recent results of Harada, Kikukawa, Kugo and 
Nakano \cite{10} on the study of the gauged NJL model to curved spacetime. In 
this way the RG-improved effective potential for the gauged NJL model in 
curved spacetime is found and its phase structure is discussed. The 
influence of the curvature to chiral symmetry breaking is described. Note
that such an approach is proved to be equivalent to the ladder Schwinger-
Dyson equations (at least in flat space) what maybe useful for
generalization of Schwinger-Dyson equations to curved spacetime.

The paper is organized as follows: In Section 2 a brief review of the 
model and the approximation used in Ref. \cite{10} is given. In Section 3 the 
gauge Higgs--Yukawa model is discussed in curved spacetime. In particular,
 the RG equation for scalar--gravitational coupling constant $\xi(t)$ and 
the one--loop effective potential are found. In Section 4, using the running 
couplings of the two previous sections and the approach of \cite{3},\cite{10} the 
RG-improved effective potential in gauged NJL model in curved spacetime 
is found (in linear curvature approximation). In the case of fixed gauge 
coupling the conditions for chiral symmetry breaking with account of the 
curvature are derived explicitly. Some remarks concerning the extension of our
results are given in the Conclusion.

\section{\large\bf Review of the class of gauge Higgs-Yukawa models}

In this section we review the class of gauge Higgs-Yukawa models in flat 
spacetime we are going to extend and the approximations we are using in 
such an analysis. Thereby we closely follow Ref. \cite{10} where, using the
 RG approach, this model has been considered in detail.

The Lagrangian of the model in flat spacetime is given by (we use the 
notations of \cite{10})

\begin{eqnarray}
L_m =&-&{\frac{1}{4}}~ G^a_{\mu\nu} G^{a\mu \nu} ~+~ {\frac{1}{2}}~
(\partial_{\mu}\sigma)^2
 ~-~\frac{1}{2}~ m^2\sigma^2~-~{\frac{\lambda}{4}}~\sigma^4 \nonumber\\
&+&{\sum\limits^{N_f}_{i=1}}~ \bar{\psi}_i ~i\hat{\D}~ \psi_i ~-~ 
{\sum\limits^{n_f}_{i=1}}~y\sigma \bar{\psi}_i\psi_i~.~
\end{eqnarray}

\noindent Here the gauge group $SU(N_c)$ is chosen, with $N_f$ 
fermions $\psi_i (i = 1, 2, ..., N_f)$ belonging to the representation 
$R$ of $SU(N_c)$, $\sigma$ is the scalar field.

Let us now describe the $1/N$ approximation for the perturbative study of 
this theory at high energies through the RG equations:\\
a) The gauge coupling constant is assumed to be small:

\begin{equation}
{\frac{g^2N_c}{4\pi}}<< 1 ~,
\end{equation}

\noindent and the RG equations are considered only in the first nontrivial
 order on $g^2$.\\
b) The number of fermions should be large enough:

\begin{equation}
N_f\sim N_c~.
\end{equation}

\noindent On the same time it is supposed that only $n_f$ fermions
 ($n_f \ll N_f$) have large Yukawa couplings, whereas the 
remaining fermions have vanishing Yukawa couplings.\\
c) The approach is assumed to be perturbative also in $1/N_c$,
so only the leading order of the $1/N_c$ expansion survives; this means 
that scalar loop contributions should be negligible

\begin{equation}
|{\frac{\lambda}{y^2}}| \le N_c~.
\end{equation}

\noindent It is interesting to note that for the minimal SM we have 
$N_c = N_f/2 = 3$ and $n_f=1$.

Within the above approximation the RG equations for the coupling constants
 in (1) are \cite{11}, \cite{12}:

\begin{eqnarray}
&{}& {\frac{dg(t)}{dt}}=-~{\frac{b}{(4\pi)^2}}~ g^3(t)~,\nonumber\\
&{}&{\frac{dy(t)}{dt}}= {\frac{y(t)}{(4\pi)^2}}~[a~y^2(t)-c~g^2(t)]~,\nonumber\\
&{}&{\frac{d\lambda (t)}{dt}} = {\frac{u~y^2(t)}{(4\pi)^2}}~[\lambda (t)-y^2 (t)]~.
\end{eqnarray}

 \noindent where
$ b=(11 N_c-4T(R)N_f)/3, c=6~C_2(R), a=u/4=2~n_fN_c$. 
For the fundamental representation we have 
$T(R) = 1/2, C_2(R) = (N_c^2 - 1)/(2N_c). t = \ln (\mu / \mu_0)$ and $\mu_0$
is the reference scale to discuss low energy physics.

The RG equation for $g^2$ (respectively $\alpha = g^2/4 \pi$) is solved by

\begin{equation}
\eta (t)\equiv {\frac{g^2(t)}{g^2_0}}\equiv {\frac{\alpha
(t)}{\alpha_0}}=
(1+{\frac{b~ \alpha_0}{2\pi}}t)^{-1}~.
\end{equation}

To solve the RG equations for the Yukawa and the scalar couplings it is 
convenient to introduce the following RG invariants:

\begin{eqnarray}
&{}& h(t)\equiv -\eta^{-1+c/b}(t)~\left[
1-{\frac{c-b}{a}} ~~
{\frac{g^2(t)}{y^2(t)}}\right],\nonumber\\
&{}& k(t)\equiv-\eta^{-1+2c/b}(t)~\left[1-{\frac{2c-b}{2a}}~~
{\frac{\lambda (t)}{y^2(t)}}~~ {\frac{g^2(t)}{y^2(t)}}\right]~.
\end{eqnarray}

\noindent With their help we obtain

\begin{eqnarray}
&{}& y^2(t)={\frac{c-b}{a}}~ g^2
(t)~\left[1+h_0\eta^{1-c/b}(t)\right]^{-1},\nonumber\\
&{}& \lambda (t) = {\frac{2a}{2c-b}}~~{\frac{y^4(t)}{g^2(t)}}~
\left[1+k_0\eta^{1-2c/b}(t)\right].
\end{eqnarray}

\noindent Of course, these solutions are explicitly known only when the
values of the RG invariants $h$ and $k$ are given.

As it has been shown in detail in Ref. \cite{10} for the solutions (6) -- (8)
to be consistent with the above approximation (2) -- (4), and in order
for the theory to be
non--trivial one , we should have (see also related discussion in
 \cite{7})

\begin{equation}
c>b~~\mbox {and}~~ 0\le h_0 < \infty ~;
\end{equation}

\noindent that makes the Yukawa coupling to be asymptotically free and 
the theory to be non--trivial (see also Ref. \cite{13}). Further analysis shows 
that the scalar coupling constant is nontrivial in the above approximation 
only if

\begin{equation}
k_0 = 0~.
\end{equation}

Below we will discuss the solutions (8) with conditions (9) and (10) only. 
The case $k_0 = h_0 = 0,~ c>b$ actually corresponds to the fixed point of 
Ref. \cite{14}. Obviously, in this case a reduction of couplings takes 
place \cite{15}, \cite{16}. However, one can get also another fixed point
 \cite{17} in 
gauged NJL model (see the discussion in \cite{3}).

\section{\large\bf Gauge Higgs-Yukawa model in curved spacetime}

Let us extend now the model of the previous section to curved spacetime. 
It is by now well--known (see Ref. \cite{18} for an introduction) that in order 
to be multiplicatively renormalizable in curved spacetime, the Lagrangian 
of the theory should be:

\begin{equation}
L=L_m+L_{ext} -{\frac{1}{2}} \xi R \sigma^2~,
\end{equation}

\noindent where $L_m$ is given by (1) with a change of flat (partial)
derivatives to the corresponding covariant 
derivatives: $\partial_\mu \rightarrow \nabla_\mu$, $\xi$ is the 
non--minimal scalar--gravitational coupling constant, and

\begin{equation}
L_{ext} =a_1 R^2+a_2 
C^2_{\mu \nu\alpha \beta} +a_3 G+a_4 \Box
R+\Lambda-{\frac{1}{\kappa}} R~.
\end{equation}

\noindent The Lagrangian $L_{ext}$ of the external gravitational field 
is necessarily introduced in order to have the theory to be multiplicatively
renormalizable one; $a_1, ... a_4, \kappa$ are gravitational coupling
constants.

Since the RG equations for those coupling constants which are present in 
flat space do not change in curved spacetime \cite{18}, all the discussions 
of Section 2 (as well as the approximations introduced there) are also valid 
here. In addition, the effective coupling constants corresponding to $\xi, 
a_1, ... , a_4, G$ appear. However, for our purposes, we only need the 
RG equation for the coupling constant $\xi$.

The technique to calculate the one--loop $\beta$-function of $\xi$ is 
quite well--known \cite{18}; therefore, we will give only the final answer 
(within the above described approximations):

\begin{equation}
{\frac{d\xi (t)}{d t}} = {\frac{1}{(4\pi)^2}}~ 2a y^2 (t)~
\left(\xi (t)- 
{\frac{1}{6}}\right)~.
\end{equation}

\noindent Taking into account that

\begin{equation}
{\frac{d}{dt}} f(t)\equiv {\frac{d}{dt}}\left\{ \eta^{c/b} (t) ~~
{\frac{(\xi (t)-{\frac{1}{6}})}{y^2 (t)}} \right\}=0
\end{equation}

\noindent we may use this RG invariant $f(t)$ to find the solution of (13):

\begin{equation}
\xi (t)={\frac{1}{6}}+y^{2}(t)\eta^{-c/b} (t)~ f_0~,
\end{equation}

\noindent where $f_0 \not= 0$ is the value of the RG invariant. 
Note that in the UV--limit ($t \rightarrow \infty$) we have:

\begin{eqnarray}
&{}& y^2 (t)\sim {\frac{c-b}{a}}~~{\frac{g_0^2}{h_0}}
 ~\eta^{c/b} (t)
\rightarrow +0~~~~~(h_0\ge 0)\nonumber\\
&{}& \xi (t) \simeq {\frac{1}{6}} + {\frac{c-b}{a}}~ g^2_0~{\frac{f_0}{h_0}}~.
\end{eqnarray}

For $f_0 \approx 0$ the scalar-gravitational coupling constant
$\xi(t) \rightarrow 1/6$, i.e. asymptotic conformal invariance is 
realized \cite{19}. Hence, the RG invariant $f_0$ characterizes the
deviation from conformal invariance ($\xi = 1/6$). Different types 
of UV--behaviour of $\xi (t) $ for different GUT's have been listed
in \cite{18}; the most typical ones are:

\begin{eqnarray}
&a)& \xi (t)\rightarrow {\frac{1}{6}}~~ \cite{18},\cite{19},\nonumber\\
&b)& |\xi (t)| \rightarrow \infty~~\cite{18},\cite{19},\nonumber\\
&c)& \xi (t) \rightarrow \xi~~ \cite{18}.
\end{eqnarray}
\noindent As we can see, unlike the analysis in the case of flat 
spacetime, there will not appear any restrictions to the sign
and the value of $f_0$ from the study of the UV--asymptotics of $\xi(t)$.

Let us discuss now the one-loop effective potential for the theory (1) 
in the approach where we keep only terms with the accuracy up to linear 
curvature \cite{20}, i.e. $\sigma^2 \gg |R|$. Using the technique of Ref.
 \cite{20} 
and working in leading order of the approximation (2) -- (4), after some 
calculations one gets:

\begin{eqnarray}
V&=&{\frac{1}{2}} m^2 \sigma^2 + {\frac{\lambda}{4}} \sigma^4 +
{\frac{1}{2}}
\xi R \sigma^2 - {\frac{a \mu^4_F}{2(4\pi)^2}} \left[ ln 
{\frac{\mu^2_F}{\mu^2}}-{\frac{3}{2}}\right]\nonumber\\
&{}& -~{\frac{a R \mu^2_F}{12(4\pi)^2}} \left[ ln
{\frac{\mu^2_F}{\mu^2}}-1\right],
\end{eqnarray}

\noindent where $\mu_F \equiv y \sigma$.

Using such a form for the potential $V$ one may investigate the 
curvature-induced phase transitions which have been discussed in detail in
Refs. \cite{18},\cite{20} for different GUT's.

\section{\large\bf Gauged Nambu-Jona Lasinio model in curved spacetime}

We now discuss the gauged NJL model in curved spacetime. This model is 
quite popular due to different reasons, for example, the
dynamical symmetry breaking in NJL models may also be relevant for the 
electroweak interaction with the top quark condensate as an order 
parameter (see \cite{2}, \cite{3}), NJL--like theories maybe used 
for equivalent descriptions of the Standard Model \cite{3}, \cite{21}.
 Moreover, the renormalizability
of gauged NJL model may be proven in some sense \cite{10} (for earlier 
discussions on renormalizability of the NJL model see \cite{22}).

Under the assumption that not elementary ones but composite bound 
states are mainly relevant in quantum cosmology (for a recent discussion 
see \cite{23}), there was an attempt to study non-gauged NJL model in curved
spacetime \cite{9},\cite{6} and, in particular, to consider the chiral symmetry 
breaking under the influence of the external gravitational field \cite{6}. 
Such an investigation has been done via explicit calculation of the 
effective potential for the composite field $<\bar{\psi}\psi>$. Our 
purpose here will be to discuss the same question on RG language for
the gauged NJL model using its equivalency with the gauged Higgs--Yukawa 
model (1) (the explicit loop calculations in such a model are extremely
hard to do).

We start from the gauged NJL model with four--fermion coupling constant G 
in curved spacetime

\begin{equation}
L=-{\frac{1}{4}} G^2_{\mu \nu} + {\sum\limits^{N_f}_{i=1}}
\bar{\psi} i \hat{\D}
\psi_i + G ~{\sum\limits^{n_f}_{i=1}} (\bar{\psi}_i \psi_i)^2~.
\end{equation}

\noindent As usually one introduces an auxiliary field $\sigma$ to replace
the NJL model by the equivalent Higgs--Yukawa model.  As is shown in 
Ref. \cite{3}, it is possible within the RG approach to use a set of boundary 
conditions for the effective couplings of the gauge Higgs-Yukawa model 
at $t_\Lambda = \ln (\Lambda/\mu_0)$ (where $\Lambda$ is the UV--cutoff
of the gauged NJL model) in order to identify the gauged NJL model with 
the gauge Higgs--Yukawa model. Taking into account Eqs. (8) the explicit
expressions for these compositeness conditions (which are specified 
in \cite{3}) 
are given as (see also Eqs. (4.7) and (4.8) of \cite{10}):

\begin{eqnarray}
&{}& y^2 (t) ={\frac{c-b}{a}}~ g^2 (t)~ \left[
1-\left({\frac{\alpha (t)}
{\alpha (t_\Lambda)}}\right)^{1-c/b} \right]^{-1}
\equiv 
y^2_\Lambda (t),\nonumber\\
&{}& {\frac{\lambda (t)}{y^4(t)}} =
{\frac{2a}{2c-b}}~~{\frac{1}{g^2 (t)}}
\left[ 1- \left( {\frac{\alpha (t)}{\alpha (t_\Lambda)}}\right)^
{1-2c/b} \right] \equiv {\frac{\lambda_\Lambda
(t)}{y^4_\Lambda(t)}}~,
\end{eqnarray}

\noindent where $ t<t_\Lambda$. Hence, the system (1) with 
cuttoff $\Lambda$ is equivalent to the system (19) with the same 
cutoff when the running coupling constants are given by (20). 
In this sense the gauged NJL model may be called a renormalizable one.
In addition, as it has been shown in \cite{10}, one should have also a
compositeness condition for the mass, whose most convenient form will 
be (for $b \rightarrow +0$; \cite{10}):

\begin{equation}
m^2 (t) = {\frac{2a}{(4\pi)^2}} \left(
{\frac{\Lambda^2}{\mu^2}}\right)^w
y^2_\Lambda (t)~\mu^2 \left[{\frac{1}{g_4(\Lambda)}} -
{\frac{1}{w}}\right]~,
\end{equation}

\noindent where $G\equiv ((4\pi)^2/a) g_4(\Lambda)/\Lambda^2 ,
 ~g_4 (\Lambda)$ is a dimensionless constant, and $w\equiv
1-\alpha/(2\alpha_C)
,~ \alpha\equiv \alpha_0$ and where $y^2_\Lambda (t)$ 
is given by the first of Eqs. (20) at $b \rightarrow +0$.

The above compositeness conditions define the gauged NJL model as 
the gauge Higgs--Yukawa model in flat spacetime. Now, since we are 
working in curved spacetime (in linear curvature approximation), 
one has to add the compositeness condition for $\xi(t_\Lambda)$. 
Making the calculations in the same way as in \cite{9} one obtains

\begin{equation}
\xi (t_\Lambda) ={\frac{1}{6}}~.
\end{equation}

Obviously, the compositeness condition for $\xi(t)$ is again the 
same one as in the non--gauged NJL model \cite{9}. Hence, the conformal 
invariance due to the renormalization group \cite{19} takes place again. 
Analysing the RG equation (13) for $ \xi(t)$ we will find that in all 
cases (i.e. for the general situation, where $y(t)$ is given by (20), 
for the case with $b \rightarrow 0$ and with the corresponding $y(t)$, 
and also for the case $\alpha_0 \rightarrow 0$, i.e. the non-gauged 
NJL model) we have

\begin{equation}
\xi (t)={\frac{1}{6}}~.
\end{equation}

\noindent In particular, for $b \rightarrow 0$ in the limit 
$\Lambda \rightarrow 0$ the fixed point of Ref. \cite{14} in curved 
spacetime is given by

\begin{equation}
y^2_\ast = {\frac{(4\pi)^2\alpha}{2a\alpha_C}},~~
{\frac{\lambda_\ast}
{y^4_\ast}}={\frac{2a}{(4\pi)^2}}~{\frac{\alpha_C}{\alpha}},~~
\xi_\ast=
{\frac{1}{6}}.
\end{equation}

Now, one can study the possibility of chiral symmetry breaking in 
the model under discussion using the effective potential language, or
more precisely,
   the RG--improved effective potential (see \cite{24},\cite{25} and
references therein). Because the technique to study the RG--improved 
effective potential is widely known for the flat space \cite{24},\cite{25}
 as 
well as for curved spacetime \cite{20} we will not give any details of its
derivation.

Using the fact that the effective potential satisfies the RG equation, 
one can explicitly solve this equation by the method of characteristics,
and we find (for more details, see \cite{24},\cite{25} and \cite{20}):

\begin{equation}
V (g,y,\lambda,m^2,\xi,\sigma,\mu)=V(g(t), y(t),\lambda (t),m^2
(t),\xi (t),
\sigma (t), \mu e^t)~,
\end{equation}

\noindent where the effective coupling constants $g(t), ... , \sigma(t)$ 
are defined by the RG equations (5), (13) (at the scale $\mu e^t$) and 
corresponding RG equations for $m^2 (t), \sigma(t)$ written in \cite{10} for 
the case of the above gauge--Higgs--Yukawa model ($t$ is left unspecified 
for the moment). As boundary condition it is convenient to use the 
one--loop effective potential (18).

In the case of the gauged NJL model one has to substitute into the 
effective potential (18) the effective couplings fulfilling compositeness 
conditions in accordance with Eqs. (25). In this way, one obtains the RG
improved effective potential in the gauged NJL model. In flat spacetime 
such calculation has been already done in Ref. \cite{10} for the case of fixed 
gauge coupling ($b \rightarrow +0$), so we may use for the running 
coupling constants (except scalar-gravitational
coupling constant) the results which are known from flat spacetime
calculations.

Taking into account that the condition

\begin{equation}
\mu e^t=\mu_F(t)
\end{equation}

\noindent serves actually for finding $t$, we get for the case of 
fixed gauge coupling

\begin{equation}
e^t=\left(\frac{\mu_F(\mu)}{\mu}\right)^{1/(2-w)}
\end{equation}

\noindent where $\mu_F(\mu) = \mu_F(t=0)$. Using (20), (23) and (27) 
in the case $b \rightarrow 0$ one gets (for the case of flat space 
see expression (6.13) of \cite{10}):

\begin{eqnarray}
{\frac{(4\pi)^2}{2a}}~{\frac{V}{\mu^4}}&=&{\frac{1}{2}}~y^2_\Lambda (\mu)
\left({\frac{\Lambda^2}{\mu^2}}\right)^w
\left({\frac{1}{g_4(\Lambda)}}-
{\frac{1}{g^\ast_4}}\right) {\frac{\sigma^2(\mu)}{\mu^2}}\nonumber\\
&{}&
+~{\frac{\alpha_C}{4\alpha}}
 ~\left[\left({\frac{y_\Lambda(\mu)\sigma(\mu)}{\mu}}\right)^{{\frac{4}{2-w}}}
 -
\left({\frac{\mu^2}{\Lambda^2}}
\right)^{{\frac{\alpha}{\alpha_C}}}
 \left({\frac{y_\Lambda(\mu)\sigma (\mu)}{\mu}}\right)^4\right]\nonumber\\
&{}& +~{\frac{3}{8}}\left({\frac{y_\Lambda (\mu)\sigma(\mu)}{\mu}}
\right)^{{\frac{4}{2-w}}} \nonumber\\
&{}&+~ {\frac{R}{24\mu^2}}  
\left({\frac{y_\Lambda(\mu)\sigma(\mu)}{\mu}} 
\right)^{{\frac{2}{2-w}}} 
\left(1+~{\frac{2\alpha_C}{\alpha}}
\right)\nonumber\\
&{}&-~
{\frac{R\alpha_C}{12\alpha \mu^2}
\left({\frac{\mu}{\Lambda}}
\right)^{{\frac{\alpha}{\alpha_C}}}~
 \frac{y^2_\Lambda (\mu)\sigma^2(\mu)}{\mu^2}}~,
\end{eqnarray}

\noindent where $g^\ast_4 \equiv w$. Thus, we have got the RG--improved 
effective potential for the gauged NJL model (with finite cutoff) in
curved spacetime.

Taking the limit $\Lambda \rightarrow \infty$ we will get the renormalized 
effective potential of the gauged NJL model in curved spacetime:

\begin{eqnarray}
{\frac{(4\pi)^2}{2a}}~ {\frac{V}{\mu^4}}&=&{\frac{1}{2}} 
\left( {\frac{1}{
g_{4R}(\mu)}}-{\frac{1}{g^\ast_{4R}}}\right){\frac{y^2_\ast~\sigma^2 (\mu)}
{\mu^2}}+\nonumber\\
&{}& +~ {\frac{\alpha_C}{4\alpha}}\left(
1+{\frac{3\alpha}{2\alpha_C}}\right)
\left({\frac{y_\ast~ \sigma(\mu)}{\mu}}\right)^{{\frac{4}{2-w}}}\nonumber\\ 
&{}&+~ {\frac{R}{24\mu^2}} \left(1+{\frac{2\alpha_C}{\alpha}}\right)
\left({\frac{y_\ast ~\sigma (\mu)}{\mu}}\right)^{{\frac{2}{2-w}}}
\end{eqnarray}

\noindent where $y_\ast$ is given by (24) and the four--fermion coupling 
renormalization has been done; in particular $g^\ast_{4R}$ is a finite 
constant (for discussion of that renormalization see \cite{10}, \cite{7}).

 Note that 
in flat space ($R=0$) the potential (29) in the same approach has been 
obtained in Ref. \cite{7} (for non-renormalized potential see also 
\cite{26}) using 
 the ladder SD equation. Hence, the RG--improved effective potential
may serve as a very useful tool to study the non--perturbative effects on 
equal footing with the SD equation. It is really surprising that RG
improved effective potential gives the same results as ladder SD equation.

Using the finite effective potential (29) we are able to discuss the 
chiral symmetry breaking in the gauged NJL model under consideration. 
In flat spacetime, the possibility of chiral symmetry breaking is defined 
by the sign of the first term in (29). That gives the value of the ciritical 
four-fermion coupling constant $g^\ast_4(\Lambda) = w$. In curved spacetime 
the situation is more complicated as one has additional terms in the
effective potential (29), even in the linear curvature approximation.

Taking into account that

\begin{eqnarray}
&{}&w=1-{\frac{3C_2(R)}{2(2\pi)^2}} g^2\simeq 1-{\frac{3}{4\pi}} 
{\frac{N_cg^2}{4\pi}} \simeq 1,\nonumber\\
&{}& {\frac{\alpha_C}{\alpha}} \simeq {\frac{2\pi}{3}}
{\frac{4\pi}
{N_c g^2}}>> 1,
\end{eqnarray}

\noindent and introducing $x=y_\ast~\sigma (\mu)/\mu$, one can 
rewrite the quadratic part of the potential (29) as follows:

\begin{eqnarray}
&{}&{\frac{(4\pi)^2}{2a}}~ {\frac{V^{(2)}}{\mu^4}}\simeq
\left\{ {\frac{1}{2}}\left(
{\frac{1}{g_{4R}(\mu)}} - {\frac{1}{g^\ast_{4R}}}\right) +
{\frac{R}{12\mu^2}}~ {\frac{\alpha_C}{\alpha}}\right\} x^2~.
\end{eqnarray}

\noindent Relation (31) determines the way to estimate the chiral 
symmetry breaking for the gauged NJL model in curved spacetime.

In particular, in flat spacetime and for (cutoff) dependent effective 
potential we observe that the chiral symmetry is broken for

\begin{equation}
{\frac{1}{g_4(\Lambda)}} - {\frac{1}{g^\ast_4}}<0 ~~\mbox{or}~~
{\frac{1}
{g_{4R}(\mu)}} - {\frac{1}{g^\ast_{4R}}}<0~.
\end{equation}

\noindent Hence, the critical value of the four-fermion coupling 
constant which divides the chiral symmetric and non-symmetric phase 
is $g^\ast_4 = w$.

 In curved spacetime, chiral symmetry is always broken if

\begin{equation}
\left( {\frac{1}{g_{4R}(\mu)}} - {\frac{1}{g^\ast_{4R}}}\right)
+
{\frac{\alpha_C}{6\alpha}}~~ {\frac{R}{\mu^2}}<0~.
\end{equation}

\noindent When this condition (33) is valid one can easily find the 
curvature--induced dynamical fermion mass.

For the cutoff--dependent
effective potential the corresponding condition looks as:

\begin{eqnarray}
&{}&\left( {\frac{\Lambda^2}{\mu^2}} \right)^w
\left({\frac{1}{g_4(\Lambda)}}-
{\frac{1}{g^\ast_4}} \right) + {\frac{R}{6\mu^2}}
{\frac{\alpha_C}
{\alpha}}\left[ 1-\left(
{\frac{\mu}{\Lambda}}\right)^{{\frac{\alpha}
{\alpha_C}}}\right] < 0~.
\end{eqnarray}

Thus, we got the condition for chiral symmetry breaking in terms of 
the dimensionless curvature $\tilde{R} \equiv R/{\mu}^2$. From (33) 
we see that the critical coupling constant depends on the curvature. 
For negative curvature the critical coupling is higher and one has a 
greater chance to find the system in the phase with broken chiral 
symmetry. As examples of spaces being in correspondence with (34) on 
can consider the inflationary universe $\sf{S_4}$ with small curvature 
or some universe with small constant curvature and a topology of the 
form $\sf{R_i \times H_{4-i}, R_i \times S_{4-i}}$, 
 ~or ~$\sf{S_i \times H_{4-i}}$, (i = 1,2,3) (see \cite{27}
for explicit examples of quantum field calculations on such backgrounds).

Note that the critical value of the curvature at which symmetry 
breaking is absent is defined according to (33) by

\begin{equation}
{\frac{R_C}{\mu^2}}={\frac{6\alpha}{\alpha_C g^\ast_{4R}}}~.
\end{equation}

\noindent At all positive curvatures below $R_c$ as well as at 
small negative curvatures the chiral symmetry is broken.

It is quite remarkable that such a simple quantum conditon of chiral 
symmetry breaking (33), (34) in gauged NJL model in curved spacetime
is obtained explicitly. Up to now such a simple symmetry breaking 
tree--level condition in curved spacetime has been known only for the 
Higgs sector of (11):

\begin{equation}
\sigma^2 = -(\xi R +m^2)/\lambda
\end{equation}

\noindent where $\xi R + m^2 < 0$.

\section{\large\bf Conclusion}

Extending the approach of papers \cite{3},\cite{10} we discussed the gauged NJL 
model in curved spacetime using quite standard RG language. In this 
way the RG--improved effective potential for gauged NJL model in curved 
spacetime is calculated. In the case of fixed gauge coupling 
($b \rightarrow +0$) the condition of chiral symmetry breaking 
is found explicitly. This condition connects the value of the 
four--fermion coupling constant with the curvature.

It is interesting to note that one can generalise the results of 
Section 4 to the case of running gauge coupling too. The only 
problem to be solved is that even in flat space the effective 
potential cannot be written explicitly, except in a few simple 
limiting cases.

To mention also another direction of consideration one can easily 
apply the above RG approach to study the gauged NJL model in a given 
specific spacetime like, for example, the DeSitter universe; this may 
be of importance for the construction of inflationary universes based 
on composite bound states.

Finally, using the RG approach in the same way one can develop the 
description of GUT's in curved spacetime via gauged NJL--like models.

SDO would like to thank K. Yamawaki and A.Wipf for useful discussions.
This work
has been supported by MEC (Spain) 
and by "Acciones Integradas Hispano--Alemanas".

\end{document}